# Influence of strain and point defects on the electronic structure and related properties of (111)NiO epitaxial films

Bhabani Prasad Sahu, Poonam Sharma, Santosh Kumar Yadav, Alok Shukla and Subhabrata Dhar*
*Department of Physics, Indian Institute of Technology Bombay, Mumbai 400076, India*
*Email: dhar@phy.iitb.ac.in

(111)NiO epitaxial films are grown on c-sapphire substrates at various growth temperatures ranging from room-temperature to 600℃ using pulsed laser deposition (PLD) technique. Two series of samples, where different laser fluences are used to ablate the target, are studied here. Films grown with higher laser fluence, are found to be embedded with Ni-clusters crystallographically aligned with the (111)NiO matrix. While the layers grown with lower laser energy density exhibit p-type conductivity specially at low growth temperatures. X-ray diffraction study shows the coexistence of biaxial compressive and tensile hydrostatic strains in these samples, which results in an expansion of the lattice primarily along the growth direction. This effective uniaxial expansion $\varepsilon_\perp$ increases with the reduction of the growth temperature. Band gap of these samples is found to decrease linearly with $\varepsilon_\perp$. This result is validated by density functional theory (DFT) calculations. Experimental findings and the theoretical study further indicate that $V_{Ni} + O_I$ and $V_O + Ni_I$ complexes exist as the dominant native defects in samples grown with Ni-deficient (low laser fluence) and Ni-rich (high laser fluence) conditions, respectively. P-type conductivity observed in the samples grown in Ni-deficient condition is more likely to be resulting from $V_{Ni} + O_I$ defects than Ni-vacancies ($V_{Ni}$).

## I. INTRODUCTION:

NiO is one of the few wide bandgap semiconductors, where stable p-type doping is achievable [1,2]. The material has a rock salt crystal structure with a bandgap that is reported to range from 3.6 to 4 eV [3–5]. Further, NiO exhibits antiferromagnetic behaviour with Neel temperature of 525K [6]. The material also shows very high chemical and thermal stability. All these properties make NiO a promising candidate for a wide range of device applications including exchange bias systems [7], spintronic devices [8], UV photodetectors and UV-light emitting diodes [9–11]. In addition, NiO is a classic example of Mott-Hubbard charge transfer insulator that makes the system interesting from basic physics point of view as well[10].

For many of the applications, epitaxial layers of NiO are preferred over polycrystalline films. There are several reports of epitaxial growth of the material on mainly c-sapphire and MgO substrates, using varied techniques, like pulsed laser deposition [13,14], mist chemical vapor deposition [3], RF magnetron sputtering [15], and molecular beam epitaxy [16]. Growth of (111)NiO layers embedded with crystallographically aligned Ni-clusters has also been reported [17]. It has been found that the shape, size and density of these clusters can be controlled by adjusting the growth parameters [19]. It is to be noted that the biaxial strain in the epitaxial layers can originate from the lattice and/or thermal expansion coefficient mismatches between the grown layer and the substrate. Another possible origin of strain could be the formation of defects/impurities in the film. Presence of these defects can cause hydrostatic strain in the film. Presence of strain in the layer can influence the band structure of the material. It is noteworthy that Ni-vacancy ($V_{Ni}$) defects are commonly believed to be the origin of unintentional p-type doping often reported in as grown NiO layers. These defects, whose formation probability depends on the Ni and oxygen flux ratio during growth, can control not only the conducting state of the layer but also the band-structural properties as a whole. It is important to mention that the large variation of the NiO band-gap values reported in the literature can be attributed to the variation of strain and defect concentration in the films. Understanding the influence of strain resulting individually from the defects and the growth on the electronic properties of the material requires a systematic strategy of experimental and band-structural studies.

There are studies, which show that the strain development along one particular crystallographic direction has the maximum influence on the band structure in certain oxide materials, such as $La_{0.7}Sr_{0.3}MnO_3$, $SrRuO_3$ and $SnO_2$ [18–20]. Chen et.al. have experimentally investigated the variation of in-plane lattice parameters due to Mg doping in (111)NiO [21]. The study reports an expansion of the lattice only in the in-plane directions as a result of Mg incorporation, which further leads to the enhancement of the band gap of the grown layer. It should be noted that the coexistence of hydrostatic and biaxial strains arising, respectively, from the defects and growth can also cause an effective unidirectional strain in the film. It is noteworthy that among the widely existing studies on NiO, detailed experimental and theoretical investigations of strain-induced changes in its band structural properties are still lacking. Therefore, in this work, we undertake a systematic investigation of the electronic structure and associated properties of this material across different strain levels, along with the presence of defects.

Here, (111)NiO epitaxial films are grown on c-sapphire substrates using pulsed laser deposition (PLD) technique at different growth temperatures ranging from room-temperature to 600℃. Two sets of samples are grown, where two different laser fluences are used for ablation. Films grown with higher laser fluence, have been found to contain Ni-clusters, which are crystallographically aligned with the (111)NiO matrix. While the layers grown with lower laser fluence and at lower growth temperatures, exhibit p-type conductivity. X-ray diffraction study shows the coexistence of biaxial compressive and tensile hydrostatic strains in all samples, which results in an expansion of the lattice primarily along the growth direction. This effective uniaxial strain along the growth direction has been found to reduce as the growth temperature increases. A linear reduction of the NiO band gap with the increase of the strain along the perpendicular direction $\varepsilon_\perp$ is found in these samples. This finding is validated by the density functional theory (DFT) calculations of the band structure. Experimental investigation along with the DFT calculations further indicate that in samples grown with Ni-deficient (low laser fluence) and Ni-rich (high laser fluence) conditions, $V_{Ni} + O_I$ and $V_O + Ni_I$ complexes are the dominant native defects, respectively. The study further suggests that the p-type conductivity observed in the samples grown in Ni-deficient condition is resulting from $V_{Ni} + O_I$ defects rather than Ni-vacancies ($V_{Ni}$).

## II. EXPERIMENTAL DETAILS

NiO films were grown on c-sapphire substrates by PLD technique. Sapphire substrates were cleaned with trichloroethylene, acetone and iso-propyl alcohol in ultrasonic bath for 5 min each before dipping them in 1:10 aqueous solution of HF in DI water for 1 min and finally rinsing in DI water. The base pressure of the chamber was maintained at less than $1 \times 10^{-5}$ mbar prior to the deposition. During the deposition, NiO pellet was ablated by an excimer laser (wavelength 248 nm, pulse width 25 ns) at a frequency of 5 Hz. While the oxygen gas with 5N purity was flown into the chamber. The depositions were carried for 25,000 laser pulses at different growth temperatures (ranging from the room temperature to 600℃) and laser fluences [2 and 3.5 J/cm$^2$] at oxygen pressure of $2 \times 10^{-2}$ mbar. Once the ablation was over for the desired number of pulses, the laser was switched off and the sample was allowed to cool naturally at the same oxygen pressure.

X-ray diffraction studies were carried out using a Rigaku SmartLab High resolution XRD system equipped with 9kW rotating Cu anode providing monochromatized Cu $K_\alpha$- radiation. ω-2θ and ω-scans were recorded in out-of-plane geometry using a vertical goniometer. While in-plane $\phi$- and $2\theta_\chi - \phi$ scans were carried out using a horizontal goniometer. Surface morphology of the samples was studied using atomic force microscopy (AFM) [Nanoscope Multimode-IV Veeco system] and field emission gun scanning electron microscopy (FEG-SEM) [JEOL JSM-7600F]. Current-voltage (I-V) measurements were carried out using Keithley 6487 Pico ammeter/voltage source. Indium contacts, which showed ohmic behaviour, were used for these measurements. Absorption spectra were recorded using a Xenon lamp as source, a 500 mm focal length monochromator (Omni-λ 500, Zolix), a photo multiplier tube (PMT), a chopper and a lock-in amplifier (Model 7265, Signal Recovery).

## III. RESULTS AND DISCUSSIONS

### A. laser fluence $F \approx 2$ J/cm$^2$

A set of samples is grown at different growth temperatures ranging from room temperature to 600℃ at a fixed oxygen pressure of $2 \times 10^{-2}$ mbar. The laser fluence is kept at ~2 J/cm$^2$ for all the samples. AFM and SEM studies show the deposition of continuous and uniform films. AFM investigation shows the roughness to be less than 2 nm for all the samples (see supplementary information S1).

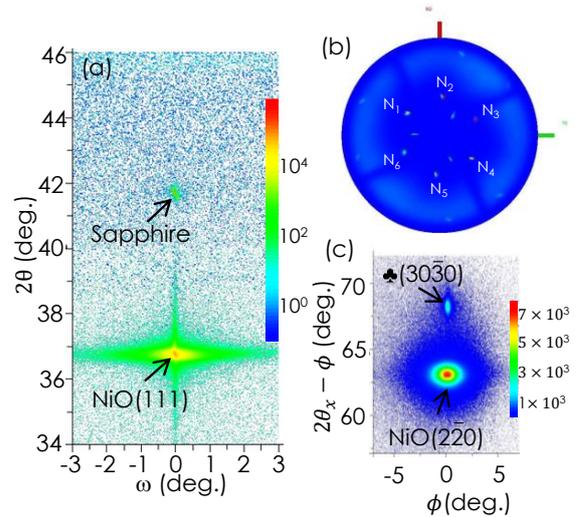

FIG. 1. (a) Reciprocal space map (RSM) in out-of-plane orientation. (b) XRD pole figure recorded for (200)NiO reflection. (c) Reciprocal space map (RSM) recorded around ($2\bar{2}0$) NiO reflection for film grown at 200℃.

Figure 1(a) shows the out of plane reciprocal space map (RSM) for the film grown at 200℃.

Reflections from (0006) sapphire and (111)NiO are clearly visible in the map. Diffraction peaks from any other asymmetric planes of NiO could not be found in the wide-angle ω-2θ profile of all the samples. This suggests [111] directional growth of NiO film in all cases. Figure 1 (b) shows the x-ray pole figure recorded on the 200°C grown film at a Bragg diffraction angle of 43.5° that corresponds to (200)NiO reflection. Pole figure plots the intensity of the diffracted beam as a function of the tilt-angle $\alpha$ ranging from 0 to 90° (represented by the radial distance) and the azimuthal angle $\beta$ ranging from 0 to 360° of the sample (represented by the azimuthal angle of the plot). Six equidistant peaks [$\Delta\beta = 60°$] at the tilt angle of $\alpha = 53.7°$ (marked by $N_1$, $N_2$, $N_3$, $N_4$, $N_5$, and $N_6$ in the figure) suggest the six-fold rotational symmetry of the basal plane. Figure 1 (c) shows the in-plane reciprocal space map (RSM) recorded around $(2\bar{2}0)$NiO peak on the film grown at 200°C. Only the peak that corresponds to $(2\bar{2}0)$NiO reflection is observed along with $(30\bar{3}0)$ reflection of sapphire. All these findings demonstrate that the NiO film is epitaxial in nature satisfying $\langle 111 \rangle$NiO$\|\langle 0001 \rangle$sapphire and $\langle 1\bar{1}0 \rangle$NiO$\|\langle 10\bar{1}0 \rangle$sapphire orientation relationships. It should be noted that above conclusion is found to be true for all the samples in this series.

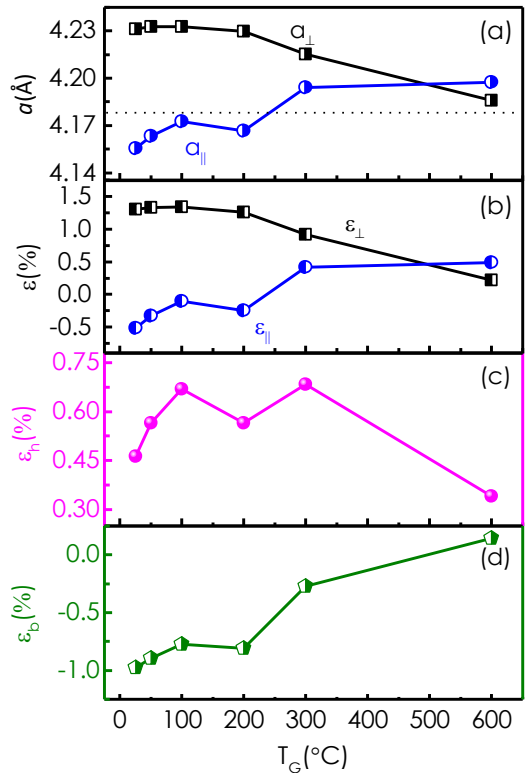

FIG. 2. (a) Lattice constants [perpendicular ($a_\perp$) and paraller ($a_\|$) to the growth direction], (b) strains [perpendicular ($\varepsilon_\perp$) and parallel ($\varepsilon_\|$) to the growth direction], (c) hydrostatic strain ($\varepsilon_h$) and (d) biaxial strain ($\varepsilon_b$) versus the growth temperature ($T_G$).

Lattice parameters in out-of-plane ($a_\perp$) and in-plane ($a_\|$) orientations of the films are calculated from the peak position of NiO (111) reflection in $\omega - 2\theta$ scan and NiO $(2\bar{2}0)$ reflection in $2\theta_\chi - \phi$ scan, respectively. Figure 2(a) plots $a_\perp$ and $a_\|$ as a function of growth temperature. It is evident that the lattice expands in out of plane direction significantly with decrease in growth temperature. Strain has been calculated using formula $\varepsilon = \frac{a-a_0}{a_0}$, where $a_0 = 4.177$Å, the lattice constant for bulk NiO. Strain in both out-of-plane ($\varepsilon_\perp$) and in in-plane ($\varepsilon_\|$) orientations are plotted versus $T_G$ in Fig. 2(b). It should be noted that the lattice mismatch between (111)NiO and c-sapphire is 7.5% [15] that should lead to an in-plane compressive biaxial strain in the NiO layer, when it is sufficiently thin. Another reason that can contribute to the strain in the NiO films specially for the samples grown at higher growth temperatures is the mismatch of the thermal expansion coefficients between NiO ($14 \times 10^{-6}$ K$^{-1}$) [22] and sapphire ($7 \times 10^{-6}$K$^{-1}$) [23], which should result in a biaxial tensile strain. As mentioned earlier, hydrostatic strain $\varepsilon_h$ resulting from the inclusion of defects/impurities, can coexists with the biaxial strain $\varepsilon_b$ in these NiO films. Contributions of the two strains can be extracted from $\varepsilon_\perp$ and $\varepsilon_\|$ data through

$$\varepsilon_h = \frac{1-\nu}{1+\nu}\left(\varepsilon_\perp + \frac{2\nu}{1-\nu}\varepsilon_\|\right) \quad (1)$$

and $\quad \varepsilon_b = (\varepsilon_\| - \varepsilon_h) \quad (2)$

where $\nu = 0.3$ is the Poisson's ratio for NiO [24,25].

In Fig. 2(c) and (d), variations of $\varepsilon_h$ and $\varepsilon_b$, respectively, with the growth temperature $T_G$ are plotted. The magnitude of $\varepsilon_b$ (compressive) clearly decreases with increasing growth temperature highlighting the role of thermal expansion coefficient mismatches for the biaxial strain generation in the layer. Evidently, $\varepsilon_h$ is always tensile in nature. $\varepsilon_h$ does not vary much up to the growth temperature of 300°C, beyond which it decreases relatively rapidly as $T_G$ is increased to 600°C.

Figure 3 plots the electrical conductivity (σ) of the films as a function of the growth temperature. Evidently, σ increases with the decrese of the growth temperarute. Conductivity in these samples can be attributed to certain native point defects in the films, which provide the background carrier concentration. Ni-defficiency induced defects, such as Ni-vacancies ($V_{Ni}$) are often believed to act as shallow acceptors that provide p-type conductivity in NiO [26,27]. This finding indicates an enhancement of Ni-deficiency induced defects in the films with the reduction of the growth temperature. Existance of tensile hydrostatic strain in the samples grown at low temperatures ($T_G \lesssim 300°C$) as shown in Fig. 2 can thus be attributed to those defects.

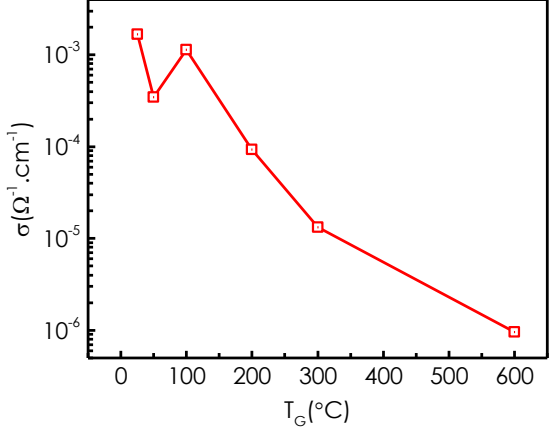

FIG. 3. Conductivity (σ) of the NiO films as a function of growth temperature ($T_G$).

B. Laser fluence $F \approx 3.5$ J/cm$^2$

Another series of samples are grown with relatively higher laser fluence of ~3.5 J/cm$^2$ at different temperatures $T_G$, while the oxygen pressure is maintained at ~$2 \times 10^{-2}$ mbar for all samples. At the increased energy density, the laser is expected to produce higher Ni-flux during growth resulting in the reduction in concentration of Ni-deficiency induced defect in the films.

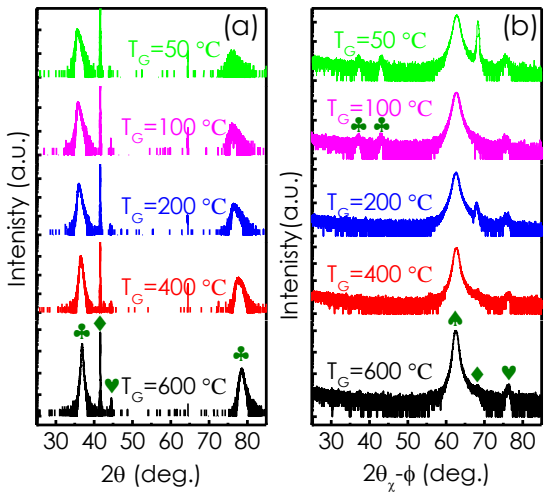

FIG. 4. (a) $\omega - 2\theta$ and (b) $2\theta\chi - \phi$ scans for samples grown at different temperatures. Symbols (♣, ♥ and ♦) represent the reflections associated with NiO, Ni and sapphire, respectively.

Figure 4(a) shows the wide angle $\omega - 2\theta$ diffractograms for samples grown at different temperatures. Observation of only (111)NiO and its higher order reflections along with (0006)Sapphire peak shows that ⟨111⟩NiO∥⟨0001⟩sapphire crystalline relationship in out-of-plane direction. Interestingly, in all samples a weak feature appears at 2θ~44.5°. This can be identified as the (111) reflection from nickel with face centred cubic (fcc) lattice structure. Fig. 4 (b) compares $2\theta_\chi - \phi$ in-plane XRD profiles for the samples. In all cases, the dominant peaks can be identified as (2$\bar{2}$0)NiO and (30$\bar{3}$0)sapphire implying the epitaxial nature of the film satisfying ⟨1$\bar{1}$0⟩NiO∥⟨10$\bar{1}$0⟩sapphire crystalline relationship in the in-plane orientation. The feature arising at $2\theta_\chi$~ 76.3° can be attributed to (2$\bar{2}$0) reflection of Ni. These results imply the inclusion of crystallographically oriented Ni-cluster in the NiO matrix. Note that similar inclusion of Ni-clusters in NiO layer has been reported by us earlier when the growth was carried out under lower oxygen pressures [17,28]. Interestingly, for the samples grown at $T_G < 100°C$, two tiny humps at $2\theta_\chi$~37.2° and ~43.5° could be seen and can be assigned to the (111) and (200) reflections of NiO, respectively. Appearance of these peaks shows the inclusion of misoriented grains in the layers grown at sufficiently low temperatures.

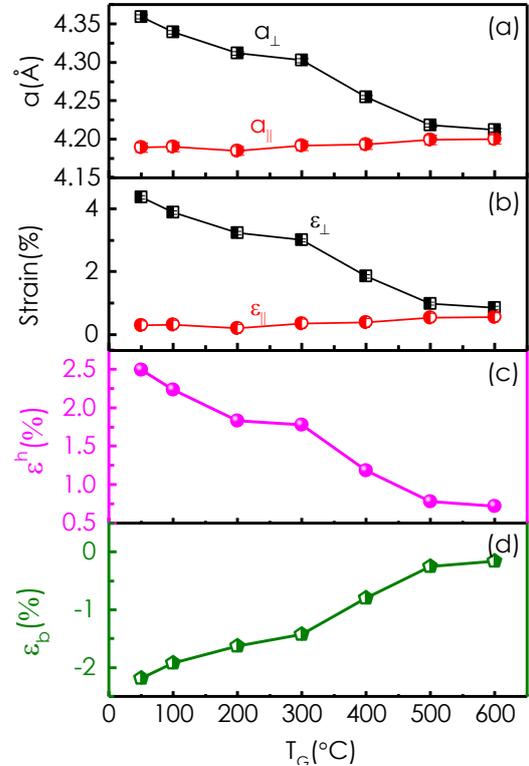

FIG. 5. (a) Lattice constants [perpendicular ($a_\perp$) and parraler ($a_\parallel$) to the growth direction], (b) strains [perpendicular ($\varepsilon_\perp$) and parraler ($\varepsilon_\parallel$) to the growth

direction], (c) hydrostatic strain ($\varepsilon_h$) and (d) biaxial strain ($\varepsilon_b$) versus the growth temperature ($T_G$).

Figure 5(a) plots $a_\perp$ and $a_\parallel$ as a function of growth temperature. It is evident from the figure that like in the previous series, the lattice mainly expands in the out-of-plane direction as the growth temperature is reduced. This is again demonstrated in Fig. 5(b), where the strains in the out-of-plane ($\varepsilon_\perp$) and in-plane ($\varepsilon_\parallel$) directions are plotted as a function of $T_G$. In Fig. 5(c) and (d), $\varepsilon_h$ and $\varepsilon_b$, respectively, are plotted versus the growth temperature $T_G$. Like in the previous series of samples, the magnitude of $\varepsilon_b$ (compressive) decreases with the increase of growth temperature, which again points to the thermal expansion coefficient mismatch between the layer and the substrate is the main reason for the generation of the biaxial strain in the films. The magnitude of the hydrostatic strain $\varepsilon_h$, which is tensile like in the previous series of samples, decreases with the rise in $T_G$.

### C. Bandgap variation

Figure 6(a) compares the absorption spectra for the samples grown at different temperatures from the series, where a lower laser fluence of $F \approx 2$ J/cm$^2$ is used for ablation. Interestingly, for samples grown at lower temperatures ($T_G \lesssim 300°C$), the band-edge is clearly redshifted as compared to the sample sample grown at $T_G = 600°C$. The band-gap does not vary much for $T_G \lesssim 300°C$. Absorption spectra for the samples grown at different temperatures from the series, where a relatively higher laser fluence of $F \approx 3.5$ J/cm$^2$ is used for ablation are presented in Fig. 6(b). The band-edge systematically shifts to the lower energy side as the growth temperature is decreased. This is evident further in the inset of the figure, where the optical gap $E_g$ obtained from these data is plotted as a function of $T_G$.

It is interesting to note that for both the series of samples it has been found that the polarity of the biaxial strain (negative) is opposite to that of the hydrostatic strain (positive). This must be the reason why the change in the lattice constant along the transverse direction ($\Delta a_\perp$) is found to be always larger than that of the horizontal direction ($\Delta a_\parallel$). This may imply that the strain in the transverse direction $\varepsilon_\perp$ must be plying an important role for the change in the band gap of NiO layer. In order to examine this further, $E_g$ is plotted versus $\varepsilon_\perp$ in Fig. 7 for all the samples investigated here. Interestingly, data for the two series of samples can be fitted with stright lines. Moreover, the slopes of the stright lines are almost the same suggesting a general tendency of the linear dependence of $E_g$ on $\varepsilon_\perp$. Note that the samples of the low laser flunce series are likely to have higher density of Ni-deficiency induced defects, which is supported by the conductivity data shown for these samples in Fig. 3. While, one can expect that the samples of the higher laser fluence series are grown in Ni-rich condition, which is supported by the presence of Ni-clusters in these samples [see Fig.4]. It is likely that the layers will be populated with oxygen deficiency induced defects. The large difference of $E_g$ between the two series of samples can be attributed to different types of defects in the two series.

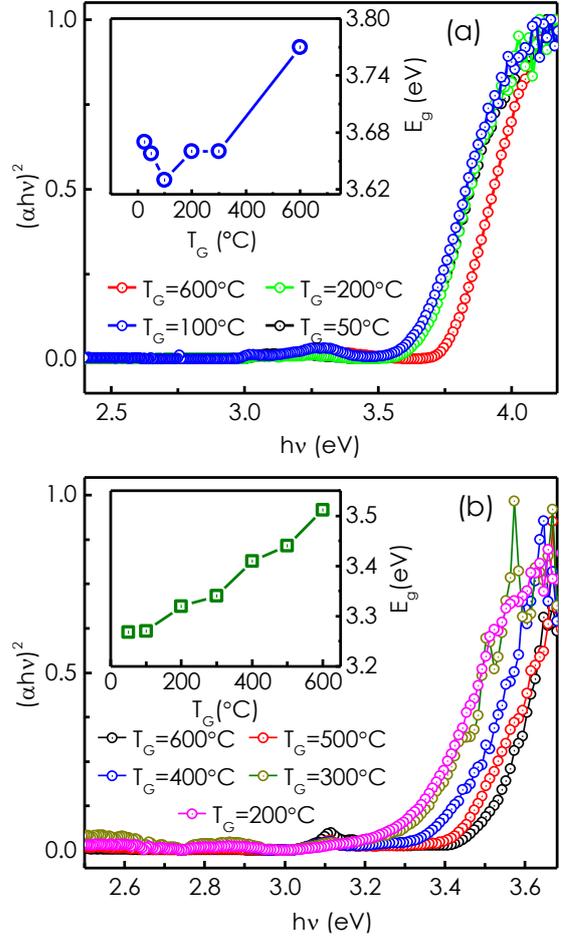

FIG. 6. The Tauc-plots for the samples grown at different temperatures in laser fluence series of (a) $F \approx 2$ J/cm2 and (b) $F \approx 3.5$ J/cm2. The respective insets show the optical bandgap ($E_g$) of the film as a function of the growth temperature $T_G$.

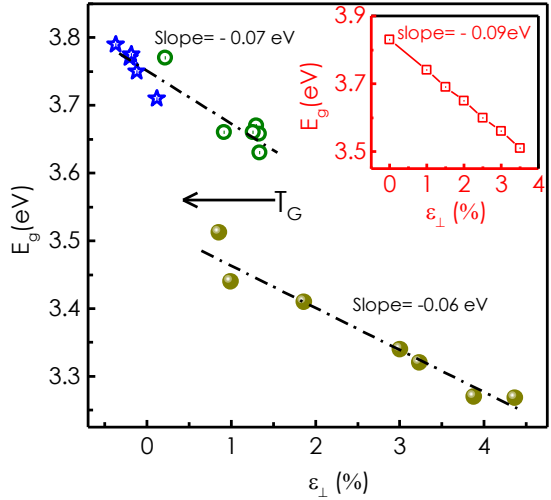

FIG. 7. Bandgap ($E_g$) as a function of $\varepsilon_\perp$. Open and solid circles represent samples from laser fluence series of (a) $F \approx 2$ J/cm$^2$ and (b) $F \approx 3.5$ J/cm$^2$, respectively. Inset shows $E_g$ versus $\varepsilon_\perp$ plot obtained from the theoretical claculations.

### D. Theoretical results

We have carried out band structure calculation within the framework of the density functional theory [29,30] by employing the Vienna *Ab-initio* Simulation Package (VASP) [31,32] to understand the role of strain and native defects on the bandgap of NiO. A $3 \times 3 \times 3$ supercell of NiO is considered for the simulations. More details of the calculation can be found in the supplementary information S2. The band structure of NiO is calculated for different amounts of tensile uniaxial strain along [111]-direction (growth direction). The band-gap ($E_g$) is plotted as a function of the strain ($\varepsilon_\perp$) in the inset of Fig. 7. It is evident from the plot that $E_g$ decreases linearly with the increase of $\varepsilon_\perp$. As shown in Fig.7, similar dependence of $E_g$ on $\varepsilon_\perp$ has been found experimentally as well. The slope of the line (-0.09 eV per % increase of strain) matches closely with that is obtaind experimentally for the series with low (-0.07 eV per % increase of strain) and high (-0.06 eV per % increase of strain) laser fluences.

Figure 8(a) and (b) compare the density of states (DOS) for NiO layer under zero strain and uniaxial tensile strain $\varepsilon_\perp$ of 2%, respectively, which clearly shows the reduction of bandgap when the uniaxial tensile strain is applied. All these observations strongly point towards the significant role of $\varepsilon_\perp$ in governing the band-gap of NiO.

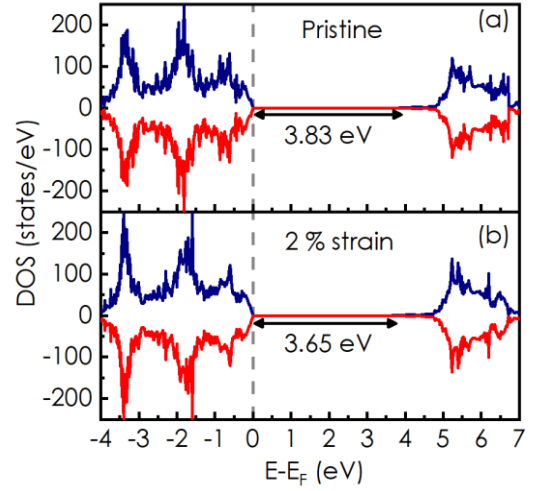

FIG. 8. Calculated total density of states a $3 \times 3 \times 3$ supercell of NiO under the out-of-plane ([111]-direction) strain of (a) $\varepsilon_\perp \sim 0$ and (b) $\varepsilon_\perp \sim 2\%$ tensile. The blue and red colors show the spin-up and spin-down states, respectively. The Fermi level is set at 0 eV.

We have extended our calculations further to explore the possibility of formation of various point defects under different growth conditions and how these defects influence the strain state of the layer. More details about the calculation of formation energy for different point defects can be seen in the supplementary information S3. Note that in each case, a solitary defect in the whole supercell is introduced for these calculations [see supplementary S4]. Figure 9(a) shows the variation of the formation energy for different point defects as a function of oxygen chemical potential in NiO. Clearly, the formation energy of Ni-vacancies ($V_{Ni}$) and O-vacancies ($V_O$) is minimum in oxygen and Ni-rich conditions, respectively. Formation of both the defect types should result in a compressive hydrostatic strain in the lattice [see supplementary table S4]. However, experimentally, in both the series of samples, the hydrostatic strain is found to be tensile in nature. This may indicate that in low (oxygen-rich growth condition) and high (Ni-rich growth condition) laser fluence series, the native defects in our NiO films are not simply $V_{Ni}$ and $V_O$. According to Fig. 9(a), the $V_{Ni}$-plus-O-interstitial ($V_{Ni} + O_I$) and $V_O$-plus-Ni-interstitial ($V_O + Ni_I$) defect complexes have the next lowest values of the formation energies in O-rich and Ni-rich conditions, respectively. From the formation energy plot we note that under the oxygen-rich conditions, it is easiest to create Ni vacancies. However, the next easiest defect to create in the O-rich conditions is Ni vacancy coupled with oxygen interstitial ($V_{Ni} + O_I$) and in the Ni-rich conditions, Oxygen vacancy coupled with Ni-interstitial ($V_O + Ni_I$), whose formation energies are smaller than those of creating

O vacancy and Ni vacancy alone in O-rich and Ni-rich conditions, respectively. It is for this reason we feel that the defect complex $V_{Ni} + O_I$ plays a very important in determining the nature of the strain in our sample. Our calculation also shows that both the defect types can give rise to hydrostatic expansion of the lattice [see supplementary S4], which is consistent with the experimental results. All these results strongly indicate that $V_{Ni} + O_I$ and $V_O + Ni_I$ type of defects dominate in our low and high laser fluence samples.

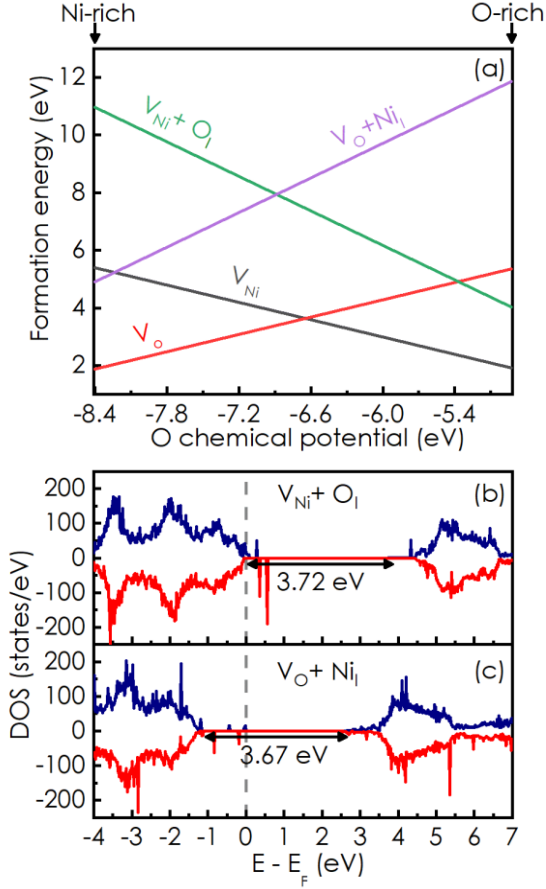

FIG. 9. (a) Calculated formation energies for different native defects in NiO. The total density of states (DOS) corresponding to (b) $V_{Ni} + O_I$ and (c) $V_O + Ni_I$ defect complexes considered in $3 \times 3 \times 3$ supercell of NiO. The Fermi level is set at 0 eV. The blue and red colors show the spin-up and spin-down states, respectively.

Figure 9(b) and (c) present the distribution of total density of states in NiO with $V_{Ni} + O_I$ and $V_O + Ni_I$ defects, respectively. Two points are important to be noticed. Firstly, the inclusion of $V_O + Ni_I$ results in a much larger reduction of band-gap of NiO [3.67 eV from 3.82 eV in pristine NiO] as compared to $V_{Ni} + O_I$ [3.72 eV], which can explain the band-gap discontinuity observed between the experimental data points associated with the two series of samples in the plot of Fig.7. Secondly, from the position of the fermi-level it is clear that holes are produced in the valence band maximum in case of NiO with $V_{Ni} + O_I$ [Fig. 9 (b)], which suggests that the defects can act as acceptors in NiO. These findings are congruous with the observation of p-type conductivity in our low laser fluence series of samples. It is noteworthy that $V_O + Ni_I$ defects do not behave as acceptors in NiO as evident from Fig. 9(c).

IV. CONCLUSION

(111)NiO epitaxial films can be grown on c-sapphire substrates within a large range of growth temperature from room temperature to 600℃ using pulsed laser deposition technique. Two sets of samples are grown with different laser fluences for ablation. In both the series of samples, biaxial compressive and tensile hydrostatic strains are found to coexist, which results in an expansion of the lattice primarily along the growth direction. This effectively uniaxial strain $\varepsilon_\perp$ has been found to decrease with the increase of growth temperature. Experimentally the study demonstrates a linear reduction of the NiO band gap with the increase of $\varepsilon_\perp$, which is validated by the band structure calculations. The study further suggests that the samples grown with Ni-deficient (low laser fluence) and Ni-rich (high laser fluence) conditions are populated with $V_{Ni} + O_I$ and $V_O + Ni_I$ defect complexes, respectively. Our investigation attributes p-type conductivity observed for the samples grown with Ni-deficient condition to $V_{Ni} + O_I$ defects, which is contrary to the common belief that Ni-vacancies ($V_{Ni}$) are responsible for unintentional background hole concentration often found in NiO.

ACKNOWLEDGEMENT

We acknowledge financial support provided by Science and Engineering Research Board (SERB) of Government of India (Grant No.CRG/2022/00l852). We would like to thank Industrial Research and Consultancy Centre (IRCC) and Centre of Excellence in Nanoelectronics (CEN), IIT Bombay for using various facilities. The computational facility (Spacetime cluster) of the Department of Physics, IIT Bombay was utilised to obtain the results of theoretical calculations.

Supplemental Materials for

# Influence of strain and point defects on the electronic structure and related properties of (111)NiO epitaxial films

Bhabani Prasad Sahu, Poonam Sharma, Santosh Kumar Yadav, Alok Shukla and Subhabrata Dhar[*]

*Department of Physics, Indian Institute of Technology Bombay, Mumbai 400076, India*

∗E-mail: dhar@phy.iitb.ac.in

## S1: Surface morphology

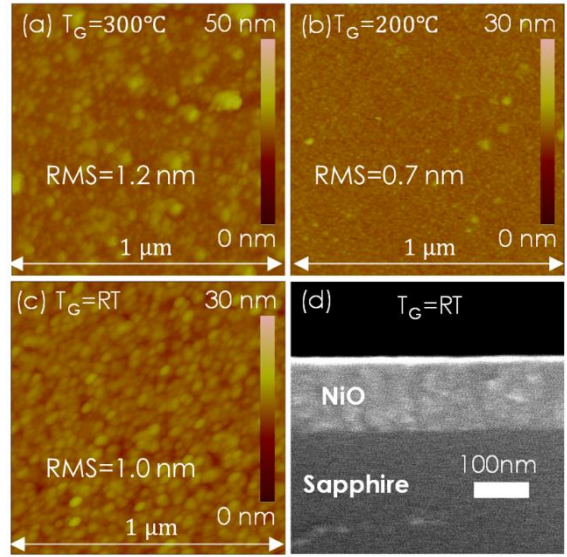

Fig. S1. (a-c) shows the AFM images of NiO films grown at different temperature for the laser fluence of $F \approx 2$ J/cm$^2$. Surface morphology of these films is uniform and smooth with RMS (root-mean-square) roughness of less than 2 nm. (d) presents the cross-sectional SEM image of the film grown at room temperature. Thickness of the films is measured to be ~150 nm.

## S2: Computational details

In order to understand the effect of strain and native defects on the electronic properties of NiO, spin-polarized density functional theory (DFT) [1,2] calculations are performed using the Vienna *Ab-initio* Simulation Package (VASP) [3,4] code with generalized gradient approximation (GGA) within Perdew-Burke-Ernzerhof (PBE) exchange-correlation functional [5]. For the study, a $3 \times 3 \times 3$ supercell of NiO (216 atoms) is taken into consideration. The plane wave energy cutoff of 500 eV and the Monkhorst-pack scheme [6] with the k-point mesh of $4 \times 4 \times 4$ is adopted for the Brillouin zone sampling. The Hellman-Feynman force on all atoms is set to $1 \times 10^{-2}$ eV, and energy threshold is fixed at $1 \times 10^{-5}$ eV for energy convergence. In order to match the band gap value with the experimental band gap, a $U$ correction on the Ni (3$d$) electrons is applied in the range of $6 \leq U(3d) \leq 11\ eV$. In our calculation, for $U = 10\ eV$, the obtained band gap is found to be in good agreement with the experimental value. Therefore, $U = 10\ eV$ is incorporated throughout the calculation.

## S3: Formation energy calculation

The formation energy ($E_{form}$) corresponding to different defect configurations is calculated using the following equation:

$$E_{form} = E_{defect} - E_{pristine} + \mu,$$

where $E_{defect}$ and $E_{pristine}$ are the total energy corresponding to defective and pristine supercells, and $\mu$ represents the chemical potential corresponding to the removed atom from the system. The chemical potential, which rely on several factors like partial pressures, growth conditions, etc., establishes the system's off-stoichiometry. At thermodynamic equilibrium, we have $\mu_{Ni} + \mu_O = \mu_{NiO}$, where $\mu_{Ni}$ and $\mu_O$ are the chemical potentials of Ni and O atoms, respectively. Further, $\mu_{NiO}$ is the energy per formula unit of bulk NiO. From our calculations, $\mu_{NiO} = -9.858$ eV is computed. Moreover, the allowed range of $\mu_{Ni}$ and $\mu_O$ is determined from their respective bulk Ni and gas $O_2$ phases. In O-rich condition, the $O_2$ gas molecule is considered as the reservoir and therefore, $\mu_O = E_{1/2\ O_2}$, where $E_{1/2\ O_2}$ is the energy per atom obtained from the $O_2$ molecule. From our DFT calculations, $\mu_O = -4.929$ eV is obtained. Therefore, using the above-mentioned relation $\mu_{Ni}$ in O-rich condition, i.e., $\mu_{Ni} = \mu_{NiO} - \mu_O = -4.921$ eV is obtained. Next, in the Ni-rich condition, bulk Ni is considered as a reservoir and, therefore, $\mu_{Ni} = E_{Ni}$ bulk, where $E_{Ni}$ shows the total energy per atom computed using the most stable bulk cubic phase of Ni with space group Fm-3m [7]. From our calculations, $\mu_{Ni} = -1.455\ eV$ is obtained, and further $\mu_O = \mu_{NiO} -$

$\mu_{Ni} = -8.403\ eV$ is computed in the Ni-rich environment. From the above discussion, the accepted range of $\mu_O$ is from -8.403 to -4.929 eV.

## S4: Variation of lattice properties with defects

For defect studies, different types of point defects i.e., Ni-vacancy ($V_{Ni}$), O-vacancy ($V_O$), Ni-vacancy plus O-interstitial ($V_{Ni} + O_i$) and O-vacancy plus Ni-interstitial ($V_O + Ni_i$) are considered in the $3 \times 3 \times 3$ supercell of NiO. In Ni-vacancy (O-vacancy) case, a single Ni (O) atom is removed from the centre position of the supercell, which corresponds to the defect concentration of 0.96 % in the system. Note that lattice parameters and supercell volume are found to decrease with inclusion of both $V_{Ni}$ and $V_O$ defects as shown in table S4. Further, $V_{Ni} + O_i$ ($V_O + Ni_i$) defect complex corresponds to the removal of one O atom and addition of one Ni at nearby interstitial site (removal of one Ni atom and addition of one O at nearby interstitial site) in the centre of the supercell. In both cases, an increment in the lattice parameters in all three directions and supercell volume are observed as shown in table S4.

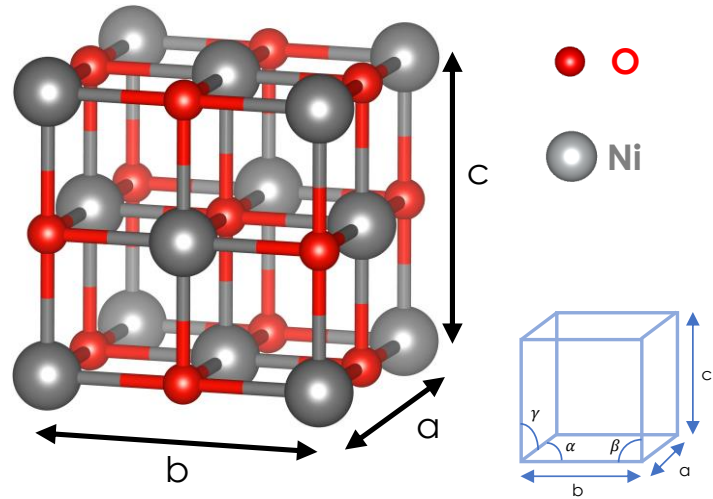

Fig. S4: Unit cell of cubic phase of NiO.

Table S4: Variation of lattice parameters, supercell volume and bandgap ($E_g$) of a $3 \times 3 \times 3$ supercell of NiO for different kinds of defect configurations.

|  | $3a$ (Å) | $3b$ (Å) | $3c$ (Å) | α (deg) | β (deg) | γ (deg) | Supercell volume (Å³) | $E_g$ (eV) |
|---|---|---|---|---|---|---|---|---|
| Pristine | 12.432 | 12.432 | 12.432 | 90 | 90 | 90 | 1921.464 | 3.83 |
| $V_{Ni} + O_i$ | 12.471 | 12.453 | 12.464 | 90 | 90 | 90 | 1935.608 | 3.71 |
| $V_O + Ni_i$ | 12.466 | 12.435 | 12.449 | 90 | 89.9 | 90 | 1931.243 | 3.65 |
| $V_O$ | 12.426 | 12.427 | 12.427 | 90 | 90 | 90 | 1918.886 | 3.92 |
| $V_{Ni}$ | 12.426 | 12.427 | 12.427 | 90 | 90 | 90 | 1918.633 | 3.88 |